\documentstyle[twoside,fleqn,espcrc2,epsf]{article}

\title{
Scalar glueball decay\thanks{Talk
presented by D.~Weingarten}}

\author{
	J.~Sexton\thanks{Permanent Address: Department of
Mathematics, Trinity College, Dublin 2, Republic of Ireland},
A.~Vaccarino and D.~Weingarten\\ 	IBM Research, P.O.~Box 218,
Yorktown Heights, NY 10598\\ }

\begin{document}

\begin{abstract}

We evaluate the coupling constant for the lightest scalar glueball to
decay to pseudoscalar meson pairs.  The calculation is done in the
valence approximation on a $16^3 \times 24$ lattice at $\beta = 5.70$
for two different values of pseudoscalar meson mass.

\end{abstract}

\maketitle

A valence approximation calculation \cite{Vaccarino} using between 25000
and 30000 gauge configurations at each of several different $\beta$
gives a value of $1740 \pm 71$ MeV for the infinite volume continuum
limit of the mass of the lightest scalar glueball.  An independent
calculation~\cite{Livertal} with between 1000 and 3000 configurations at
several different $\beta$, when interpreted~\cite{Weingarten94}
following Ref.~\cite{Vaccarino}, yields $1625 \pm 94$ MeV for the
lightest scalar glueball mass.  Finally, an analysis in
Ref.~\cite{Weingarten94} suggests that the error in the valence
approximation to the scalar glueball mass is probably less than 60 MeV.
The mass calculation with larger statistics then appears to favor
$f_J(1710)$ as a scalar glueball, with $f_0(1590)$ a possible but less
likely candidate. A third alternative is that the scalar glueball is a
linear combination of the two states with significant contributions from
each.  No other well established resonance consistent with scalar
glueball quantum numbers lies within 200 MeV of the 1740 MeV prediction.

The interpretation of the scalar glueball as $f_J(1710)$ or some
combination of $f_J(1710)$ and $f_0(1590)$ is obscured, however, by two
unresolved issues. First, if the scalar glueball decay width were large
enough, this particle would not be easily identified, in which case it
need not be any well established resonance. Second, since the scalar
glueball is a flavor singlet, one might expect its coupling constant to
a pair of $\eta's$ to be the same as its coupling to a pair of $K's$ or
a pair of $\pi's$. In fact, the observed values of these three coupling
constants are far from equal both in the case of $f_J(1710)$ and in the
case of $f_0(1590)$.

We are now in the process of calculating the scalar glueball's coupling
to pairs of pseudoscalars.  Our preliminary results strongly suggest
this particle has a width significantly below 200 MeV. Thus we consider
it quite likely that the scalar glueball has already been found in
experiment.  In addition, the coupling constants we obtain show a
variation with the mass of the decay products consistent with the
variation in both the $f_J(1710)$ and the $f_0(1590)$ couplings.

To evaluate the glueball decay couplings, consider a euclidean lattice
gauge theory, on a lattice $L^3 \times T$, with the plaquette action for
the gauge field, the Wilson action for quarks, and the gauge field
transformed to lattice Coulomb gauge.  Let $U^r_i(x)$ for a space
direction $i = 1, 2, 3,$ and integer radius $r \ge 1$, be a smeared link
field found by averaging the $r^2$ links running in direction $i$ from
the sites of the $(r-1) \times (r-1)$ square oriented in the two
positive space directions orthogonal to $i$ starting at site $x$. Let
$V^r_{ij}(x)$ be the trace of the product of $U^r_i(y)$ and $U^r_j(y)$
around the boundary of an $(r-1) \times (r-1)$ square in the $i j$ plane
beginning at $x$. Let the zero-momentum scalar glueball operator
$g^r(t)$ be the sum of the $V^r_{ij}(x)$ for all $i, j$ and $x$ in the
time $t$ lattice hyperplane. Define the quark and antiquark fields
$\overline{\Psi}^{r'}(x)$ and $\Psi^{r'}(x)$ to be Wilson quark and
antiquark fields smeared by convoluting the local Wilson fields with a
space direction gaussian with root-mean-square radius $r'$. Define the
smeared pseudoscalar field $\pi^{r'}_i(x)$ with flavor index $i$ to be
$\overline{\Psi}^{r'}(x) \gamma^5
\lambda_i \Psi^{r'}(x)$, where $\lambda_i$ is a Gell-Mann flavor matrix.
Let $\tilde{\pi}^{r'}_i(\vec{k},t)$ be the Fourier transform of
$\pi^{r'}_i(x)$ on the time $t$ lattice hyperplane.  We now fix $r$ and
$r'$ to the values $3$ and $\sqrt{6}$, respectively, used in the summary
of results presented below, and omit the radius superscripts.

At large time separations $t$, the vacuum expectation values $<
\tilde{\pi}^{\dagger}_i(0,t) \tilde{\pi}_i(0,0)>$ and $<
\tilde{\pi}^{\dagger}_i(\vec{k},t) \tilde{\pi}_i(\vec{k},0)>$ approach
$\eta_1^2 L^3 exp( -E_{\pi 1} t)$ and $\eta_2^2 L^3 exp( -E_{\pi 2} t)$,
respectively. Here $\vec{k}$ is any momentum with norm $2 \pi L^{-1}$,
$E_{\pi 1}$ is the mass of a pion, and $E_{\pi 2}$ is the energy of a
pion with momentum $2 \pi L^{-1} $. In addition, in the valence
approximation, $<g(t) g(0))>$ approaches $\eta_g^2 L^3 exp( -E_g t)$,
where $E_g$ is the mass of the lightest scalar glueball.  At this point
it is convenient to redefine $g(t)$, $\tilde{\pi}_i(0,t)$ and
$\tilde{\pi}_i(\vec{k},t)$ by dividing each by the corresponding $\eta$.

Define the two pion zero momentum operator $\tilde{\Pi}( 0, t)$ to be
$6^{-1/2} \sum_i \tilde{\pi}_i(0,t) \tilde{\pi}_i(0,0)$.  Let the two
pion operator $\tilde{\Pi}( k, t)$ for nonzero $k$ be $6^{-1} \sum_{i
\vec{k}} \tilde{\pi}_i(\vec{k},t) \tilde{\pi}_i(-\vec{k},0)$ where the
sum is over the six ways of placing a vector $\vec{k}$ of length $k$ in
a positive or negative lattice coordinate direction.  Define the vacumm
subtracted operators $\tilde{\Pi}_s( k, t)$ to be $\tilde{\Pi}( k, t) -
<\Omega| \tilde{\Pi}( k, t)|\Omega>$.  Let $| 1>$ be the lowest energy
eigenstate in $\tilde{\Pi}_s( 0, t) |\Omega>$, and let $| 2>$ be the
lowest energy eigenstate in $\tilde{\Pi}_s( 2 \pi L^{-1}, t) |\Omega>$.
The states $| 1>$ and $| 2>$ are both normalized to 1.  Let
$E_{\pi\pi1}$ and $E_{\pi\pi2}$ be the energies of these two states,
respectively.  Define the amplitudes $\eta_{i j}(t)$
\begin{eqnarray}
\label{defamp1}
\eta_{i 1}(t) & = & L^{-3} < i| \tilde{\Pi}_s( 0, t) | \Omega >, \\
\label{defamp2}
\eta_{i 2}(t) & = & L^{-3} < i| \tilde{\Pi}_s( 2\pi L^{-1}, t) | \Omega >,
\end{eqnarray}
For large $t$, $\eta_{i j}(t)$ has the asymptotic form
$\eta_{i j} exp(-E_{\pi j} t)$.
Connected three-point functions from which coupling constants can be
extracted are now given by
\begin{eqnarray}
\label{deft1}
T_1( t_g, t_{\pi})  & = & < g(t_g) \tilde{\Pi}_s( 0, t_{\pi}) >, \\
\label{deft2}
T_2( t_g, t_{\pi})  & = & < g(t_g) \tilde{\Pi}_s( 2\pi L^{-1}, t_{\pi}) >.
\end{eqnarray}

If the quark mass, and thus the pion mass, is chosen so that $E_{\pi\pi
1}$ is equal to $E_g$, the lightest intermediate state which can appear
between the glueball and pions in a transfer matrix expression for $T_1(
t_g, t_{\pi})$ is $| 1>$. Thus for large enough $t_g$ with $t_{\pi}$
fixed, $T_1( t_g, t_{\pi})$ will be proportional to the coupling contant
of a glueball to two pions at rest.  If the quark mass is chosen so that
$E_{\pi\pi 2}$ is equal to $E_g$, however, the lightest intermediate
state which can appear between the glueball and pions in a transfer
matrix expression for $T_2( t_g, t_{\pi})$ is still $| 1>$, not $| 2>$.
This holds since $\eta_{1 2}(t)$ in Eq.~(\ref{defamp2}) has no reason to
vanish.  To obtain from $T_2( t_g, t_{\pi})$ the coupling of a glueball
to two pions with momentum $2 \pi L^{-1}$, the contribution to $T_2(
t_g, t_{\pi})$ arising from the $| 1>$ intermediate state must be
cancelled off. From the three-point functions defined in
Eqs.~(\ref{deft1}) and (\ref{deft2}) we therefore define the amplitudes
\begin{eqnarray}
\label{deft1s}
S_1( t_g, t_{\pi})  = T_1( t_g, t_{\pi})  -
\frac{\eta_{21}(t_{\pi})} {\eta_{22}(t_{\pi})} T_2( t_g, t_{\pi}), \\
\label{deft2s}
S_2( t_g, t_{\pi})  = T_2( t_g, t_{\pi})  -
\frac{\eta_{12}(t_{\pi})}{\eta_{11}(t_{\pi})} T_1( t_g, t_{\pi}).
\end{eqnarray}
In $S_2( t_g, t_{\pi})$ the contribution of the undesirable $| 1>$
intermediate state has been cancelled.  In
$S_1( t_g, t_{\pi})$ a
contribution from the intermediate state $| 2>$ has been cancelled.
Although the subtraction in
$S_1( t_g, t_{\pi})$
is irrelevant for large enough $t_g$, we
expect that as a result of this subtraction $S_1( t_g,t_{\pi})$
will approach its large $t_g$ behavior more
rapidly than does $T_1( t_g, t_{\pi})$.

An additional intermediate state which can also appear in a transfer
matrix expression for either $T_i( t_g, t_{\pi})$ is the isosinglet
scalar bound state of a quark and an antiquark.  The relative
contribution of this state to either three-point function, however, is
suppressed by the factor $\mu |t_g - t_{\pi}|$, where $\mu$ is the
coupling constant of this state to the glueball. Phenomenological
arguments following Ref.~\cite{Weingarten94} strongly suggest that $\mu$
is less than 60 MeV. In our main results below, removing the additional
intermediate state would therefore give at most a 4\% correction.

At large
$t_g$ and $t_{\pi}$, the three-point functions become
\begin{eqnarray}
\label{S1asym}
\lefteqn{S_1( t_g, t_{\pi}) \rightarrow}
\nonumber \\
& & \frac{\sqrt{3} \lambda_1 \eta_{11}(1 - r_{12}r_{21}) L^3}
{\sqrt{16E_g E_{\pi 1}^2}}
s_1(t_g, t_{\pi}), \\
\label{S2asym}
\lefteqn{S_2( t_g, t_{\pi}) \rightarrow}
\nonumber \\
& & \frac{3 \lambda_2 \eta_{22} (1 - r_{12}r_{21}) L^3 }
{\sqrt{8E_g E_{\pi 2}^2}}
s_2(t_g, t_{\pi}).
\end{eqnarray}
Here $\lambda_1$ and $\lambda_2$ are the glueball coupling constants to
pseudoscalar pairs at rest or with momentum $2 \pi L^{-1}$,
respectively.  Up to a factor of $-i$ these coupling constants are
invariant decay amplitudes with the standard normalization convention.
The factors $\eta_{ij}$ are given by the large $t$ behavior of
$\eta_{ij}(t)$ as discussed earlier, and $r_{ij}$ is
$\eta_{ij}/\eta_{jj}$.  The factors $s_i(t_g, t_{\pi})$ are
\begin{eqnarray}
\label{defsi}
\lefteqn{s_i(t_g, t_{\pi}) =
\sum_t exp( -E_g |t - t_g| -} \nonumber \\
& & E_{\pi i}|t - t_{\pi}| - E_{\pi i}|t|).
\end{eqnarray}
In Eq.~(\ref{defsi}) we make the simplifying assumption that
$E_{\pi\pi i}$ is $2 E_{\pi i}$.
In our results to be presented below we actually
use a slightly more complicated expression for $s_i(t_g, t_{\pi})$
which takes into account small differences between
$E_{\pi\pi i}$ and $2 E_{\pi i}$.

To obtain values of $\lambda_i$ from Eqs.~(\ref{S1asym}) and
(\ref{S2asym}) we need the amplitudes $\eta_{ij}(t)$. These we determine
from propagators for two pion states.  Define the two pion operator
$\Pi(t_1, t_2)$ from smeared pion fields, not Fourier transformed and
not renormalized, to be $6^{-1/2}\sum_i \pi_i(0,t_1) \pi_i(0,t_2)$.
Define two pion propagators by
\begin{eqnarray}
\label{defC1}
\lefteqn{C_1(t_1,t_2) =} \nonumber \\
& & < \Pi(t_1 + 2 t_2, t_1 + t_2) \tilde{\Pi}_s( 0, t_2)>, \\
\label{defC2}
\lefteqn{C_2(t_1,t_2) =} \nonumber \\
& &  < \Pi(t_1 + 2 t_2, t_1 + t_2) \tilde{\Pi}_s(2 \pi/L, t_2)>.
\end{eqnarray}
For large values of $t_1$, these amplitudes approach
\begin{eqnarray}
\label{Casym}
\lefteqn{C_i(t_1,t_2) = } \nonumber \\
& & C_{1i} exp( -E_{\pi\pi 1} t_1) + C_{2i} exp( -E_{\pi\pi 2} t_1), \\
\label{Cxasym}
\lefteqn{C_{ij} =} \nonumber \\
& & \eta_1^2 \eta_{ij}(t_2) \eta_{i1}(t_2) + \sqrt{6}
\eta_2^2 \eta_{ij}(t_2) \eta_{i2}(t_2),
\end{eqnarray}
where the $\eta_i$ are determined from the asymptotic behavior of single
pion propagators as discussed earlier.

At $\beta$ of 5.7, we found that $k$ of 0.1650 nearly gives $E_{\pi\pi
1}$ equal to $E_g$, and $k$ of 0.1675 nearly gives $E_{\pi\pi 2}$ equal
to $E_g$.  For $k$ of 0.1650 and $t$ from 0 to 5, we determined
$\eta_{ij}(t)$ with Eqs.~(\ref{Casym}) and (\ref{Cxasym}) applied to
$C_i(t_1, t_2)$ from an ensemble of 100 independent gauge configurations
on a $16^3 \times 40$ lattice. For $k$ of 0.1675 and $t$ of 0 to 5, we
determined $\eta_{ij}(t)$ from $C_i(t_1, t_2)$ using
870 independent gauge configuratons.  The $\eta_{ij}(t)$ turn out to be
nearly independent of $t$. In each case $\eta_{00}$ is statistically
consistent with 1.0, $\eta_{11}$ is between 1.0 and 1.1 and $\eta_{12}$
and $-\eta_{21}$ are positive and less than 0.1.

\begin{figure}
\begin{center}
\vskip -5mm
\leavevmode
\epsfxsize=65mm
\epsfbox{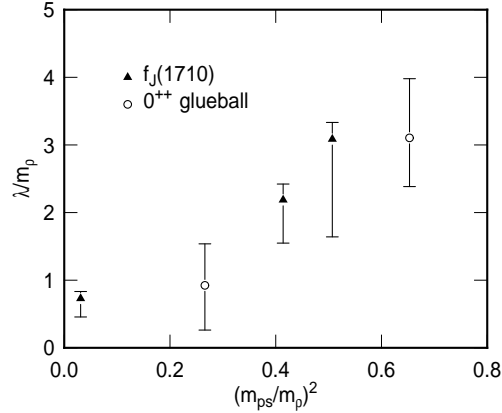}
\vskip -13mm
\end{center}
\caption{
Decay couplings.}
\label{fig:lambdas}
\vskip -8mm
\end{figure}

We then evaluated $\lambda_1$ for $k$ of 0.1650 and $\lambda_2$ for $k$
of 0.1675 using Eqs.~(\ref{S1asym}) and (\ref{S2asym}) applied to $S_1(
t_g, t_{\pi})$ and $S_2( t_g, t_{\pi})$ found with an ensemble of 7200
independent gauge configurations on a lattice $16^3 \times 24$. We
obtained statistically significant results for $t_g - t_{\pi}$ of 0, 1
and 2 and $t_{\pi}$ of 1, 2, 3, 4, and 5.  The $\lambda_i$ are nearly
constant, within statistical errors, across these 15 points.
Figure~\ref{fig:lambdas} shows predicted coupling constants fitted to
the data with $t_g - t_{\pi}$ of 1 in comparision to observed decay
couplings for decays of $f_J(1710)$ to pairs of $\eta's$, $K's$ and
$\pi's$.  The horizontal axis gives the squared mass of the decay
products. Masses and decay constants are shown in units of the $\rho$
mass. If couplings to $\eta's$, $K's$ and $\pi's$ are determined from
our predicted couplings by linear interpolation in the squared mass of
the decay products, we predict a total width for glueball decay to
pseudoscalar pairs of $52^{+45}_{-14}$ MeV, in comparison to $84 \pm 23$
MeV for $f_J(1710)$.

We would like to thank Steve Sharpe for finding an error in an
earlier version of this work.

\begin{thebibliography}{9}

\bibitem{Vaccarino} H.\ Chen, et al.,
Nucl.\ Phys.\ B (Proc.\ Suppl.) 34 (1994) 357.
\bibitem{Livertal} G.\ Bali et al., Phys.\ Lett.\ B 309 (1993) 378.
\bibitem{Weingarten94} D.\ Weingarten, Nucl.\ Phys.\ B (Proc.\ Suppl.)
34 (1994) 29.

\end{thebibliography}
\end{document}